\documentclass[conference]{IEEEtran}
\usepackage{color}

\usepackage{graphicx}
\usepackage{multirow}
\usepackage{booktabs}
\usepackage[table]{xcolor}
\begin{document}

\title{CEEC: Centralized Energy Efficient Clustering\\ A New Routing Protocol for WSNs}

\author{\IEEEauthorblockN{M. Aslam, T. Shah, N. Javaid, A. Rahim, Z. Rahman, Z. A. Khan$^{\S}$\\}
        Department of Electrical Engineering, COMSATS\\
        Institute of Information Technology, Islamabad, Pakistan. \\
        $^{\S}$Faculty of Engineering, Dalhousie University, Halifax, Canada.
        }

\maketitle

\IEEEpeerreviewmaketitle

\section{Introduction}
Energy efficient routing protocol for Wireless Sensor Networks (WSNs) is one of the most challenging task for
researcher. Hierarchical routing protocols have been proved more energy efficient routing protocols, as compare to flat and location based routing protocols. Heterogeneity of nodes with respect to their energy level, has also added extra lifespan for sensor network. In this paper, we propose a Centralized Energy Efficient Clustering (CEEC) routing protocol. We design the CEEC for three level heterogeneous network. CEEC can also be implemented in multi-level heterogeneity of networks. For initial practical, we design and analyze CEEC for three level advance heterogeneous network. In CEEC, whole network area is divided into three equal regions, in which nodes with same energy are spread in same region.

\section{Problem Statement and Objective}
In previous research work, SEP, E-SEP, and DEEC are designed for heterogeneous networks [2-4]. But these protocols do not provide any network deployment planning. Because of this, nodes with extra energy (advance nodes which have to become cluster-heads more frequently) are not uniformly dispersed throughout the network. Furthermore, these protocols use distributed clustering algorithm that increase computational overhead on all nodes. An other
problem is that, optimum number of cluster-heads are also not guaranteed through distributed algorithm. We propose CEEC routing protocol to address these issues. In CEEC, Base Station (BS) centrally selects optimum number of cluster-heads. CEEC enhances the stability and network lifetime. We simulate our proposed routing protocol using MATLAB. The results of simulations verify that our proposed model provide better network life time as compare to LEACH, SEP, E-SEP and DEEC. Next section describes the CEEC's network heterogenous network model for our proposed protocol.

\section{CEEC's Advanced Heterogeneous Model}

In WSNs, nodes are randomly dispersed in network area without any deployment management. Although nodes deployment
is very challenging task in WSNs, however, we can still address this issue by dividing whole network area into multiple logical
regions. We present an advance heterogeneous network model in this section. Our proposed network model contains three
different types of nodes called, normal, advance and super nodes. These nodes preserve different levels of energy.
We divide whole network's M$\times$M area into three equal rectangular regions Low Energy Region (LER), Medium Energy
Region (MER), and Higher Energy Region (HER). We assume that BS is placed at top of the network. We homogeneously spread normal nodes in nearest region of LER with respect to BS. Advance and Super nodes are homogeneously placed in MER and HER region respectively. Overall heterogeneous network is produced by combining allregions, as shown in Fig. 1.

\vspace{-0.3cm}
\begin{figure}[h]
\centering
\includegraphics[scale=0.5]{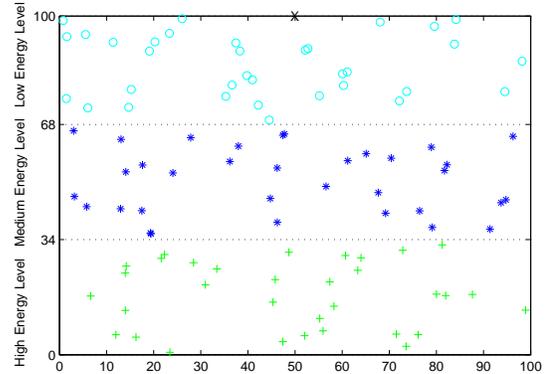}
\vspace{-0.5cm}
\caption{Distribute Heterogeneous Network Model}\label{abc}
\end{figure}
\vspace{-0.3cm}

One more distinguish feature of our proposed heterogeneous network model is that, nodes associate
with their own type of cluster-heads nodes as shown in Fig. 2.

\vspace{-0.3cm}
\begin{figure}[h]
\centering
\includegraphics[scale=0.5]{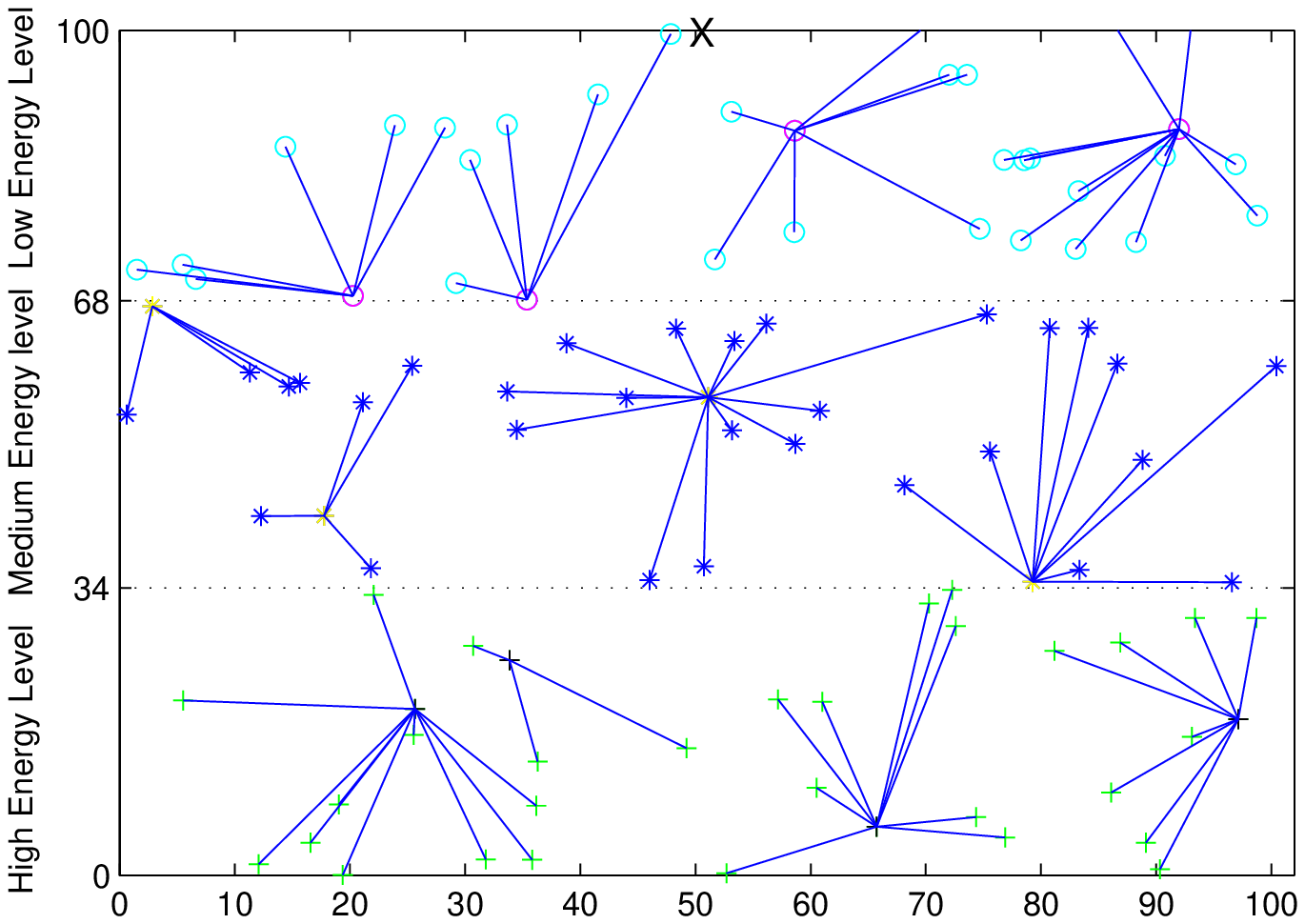}
\vspace{-0.5cm}
\caption{Distribute Heterogeneous Network Model}\label{abc}
\end{figure}
\vspace{-0.5cm}

Total $N$ nodes are scattered in whole network.

\vspace{-0.3cm}
\begin{equation}\label{}
    E_{n}=N1+N2+N3
\end{equation}

$N1$ are normal nodes, $N2$ are advance nodes and
$N3$ are super nodes. In three level heterogeneous network,
energy assigned to normal nodes is $E_0$. Advance and super
nodes have $\alpha$  and $2\times \alpha$ factors more energy respectively as
compared to normal nodes. Total energy of all normal nodes is:

\begin{equation}\label{}
    E_n=N1\times E_0
\end{equation}

Total energy of advance nodes is:

\begin{equation}\label{}
   E_a=N2\times (E_0+(1\times \alpha ))
\end{equation}

Similarly, total energy of super nodes can be calculated as:

\begin{equation}\label{}
E_{s}=N3\times (E_0+(1\times 2\alpha ))
\end{equation}

In this way, total energy of all nodes is:

\begin{eqnarray}\nonumber
E_{t}=N1\times E_{O}+N2\times (E_{O}+(1\times \alpha )) \\ +N3\times (E_{O}+(1\times 2\alpha ))
\end{eqnarray}

From above equations, it is clearly understandable that proposed advance heterogeneous network model spread the nodes in network area with the ascending order of energy. As the distance of nodes from BS increases, energy level of the nodes is also increases. It brings the equal distribution of resources with respect to responsibilities of nodes.

\section{Proposed CEEC}

In current section, we propose a Centralized Energy Efficient Clustering (CEEC) routing protocol. In earlier section, we propose advance heterogeneous network model, in which nodes with different energy level are deployed in separate regions. In CEEC, BS performs central clustering formation in network, with help of central control algorithm of CEEC. Advance central control algorithm considers four factors for selection of cluster-heads, initial energy of nodes, residual energy of nodes, average energy of each region and location of nodes. Operation of CEEC is based on rounds, with adjustable duration. Each round is divided into Network Settling Time (NST) and Network Transmission Time (NTT). During NST cluster-heads are selected and multiple clusters are formed. During NTT, sensed information from all nodes is transmitted to BS with help of cluster-heads.

\subsection{Network Settling Time (NST)}
Efficient cluster formation is key technique to enhance the network lifetime. During NST suitable cluster-heads are selected by BS, with the help of central control algorithm. In central control algorithm, BS calculates three different average energies for normal, advance and super nodes to
obtain separate cluster-heads for all regions. BS knows the initial energy of all nodes for the first round and it can simply calculate the average energies for first round. After first round, nodes provide their residual energy information to BS. Another significance of our proposed protocol is that nodes provide their residual energy information along data packets transmitted in NTT. This factor also saves their extra transmission energy cost, paid during NST, as it is paid in conventional centralized control protocols. Average energy of residual energy of all normal nodes, which are spread in closest LER to BS, is calculated by:

\begin{equation}\label{}
    \overline{E_{n}(r)}=\frac{1}{N1}\sum_{i=1}^{N1}E_{(ni)}(r)
\end{equation}

Where, $\overline{E_{n}(r)}$ is average energy and r is current round of
operation. Similarly average energy of advance nodes, which
are spread in MER to BS, is calculated by:

\begin{equation}\label{}
    \overline{E_{a}(r)}=\frac{1}{N2}\sum_{i=1}^{N2}E_{(ai)}(r)
\end{equation}

Where, $N2$ are advance type of nodes.
Average energy of super nodes, which are spread in HER to
BS, will be:

\begin{equation}\label{}
    \overline{E_{s}(r)}=\frac{1}{N3}\sum_{i=1}^{N3}E_{(si)}(r)
\end{equation}

Where, $N3$ are super nodes and $\overline{E_{s}(r)}$ is their average energy.

After calculation of average energy of each region, BS compares energy of each node to their corresponding region's average energy. Nodes with higher or equal energy to average energies $(E_{i}\geq Average Energy)$ are selected by BS as Expected Cluster-Heads (ECHs). BS has to select desired percentage $P$ of cluster-heads in every round, for each type of nodes. If number of ECHs are greater than required CHs, BS will select $AliveNodes\times P$ cluster-heads with maximum residual energy and minimum distance to BS. These finally elected cluster-heads will be grouped as Finally Selected
Cluster-Heads (FSCHs).

BS multi-casts announcements of selection to FSCHs, instead of broadcasting to all nodes, as it happens in previous centralized routing protocols. It also reduces computational over-head of non-cluster-head nodes. FSCHs receive the final decision of selection from BS and advertise their status updates to
all nodes laying in their range. If non-cluster-head node receives multiple advertisements, then it selects its clusterhead with high Received Signal Strength Indication (RSSI) and link quality. Non-cluster-head nodes send their association request to their Corresponding Cluster-Heads (CCHs) using CSMA-CA. The main restriction in association of nodes is that, these nodes have to select cluster-head of their own region. Then CCHs assign specific TDMA slots to its member nodes for data transmission during NTT. NST is very small as compare to NTT and total duration of single NST is between the end of a NTT to start of next NTT.

\subsection{Network Transmission Time (NTT)}

NTT is similar to LEACH and other clustering routing protocols. In NTT all nodes send their data to their CCHs, in assigned time slots. Cluster-heads receive the data from its cluster and aggregate the data. Data aggregation is key technique in order to compress data amount. Cluster-heads
only send meaningful information to BS in order to prolong the battery lifetime.

\section{Simulation Results for Performace of CEEC }
We simulate CEEC along with LEACH, SEP, E-SEP and DEEC to judge the performance of our proposed protocol. Simulation parameters are given table 1.

\begin{table}[h]
  \centering
  \caption{Simulation Parameters}
    \begin{tabular}{rrrrrrrrrr|rr}
    \toprule
    Parameter                                                           &       &       &       & value &  \\ \hline
    \midrule
    Network size        &       &       &       &            100m * 100m      &  \\\hline
    Initial Energy        &       &       &       &                .5 j      &  \\\hline
    P       &       &       &       &                      .1 j      &  \\\hline
    Data Aggregation Energy cost      &       &       &       &                      50pj/bit j      &  \\\hline
    Number of nodes       &       &       &       &                     100      &  \\\hline
    Packet size        &       &       &       &             200 bit       &  \\\hline
    Transmitter Electronics (EelectTx) &       &       &       & 50 nj/bit      &  \\\hline
    Receiver Electronics (EelecRx)     &       &       &       & 50 nj/bit &  \\\hline
    Transmit amplifier (Eamp)            &       &       &       & 100 pj/bit/m2 &  \\\hline

    \bottomrule
    \end{tabular}%
  \label{tab:addlabel}%
\end{table}%

Fig. 3 shows stability period of network. CEEC is about $40 \%$, $70\%$, $70\%$, $100\%$ better in stability as compare to DEEC, SEP, E-SEP and LECAH respectively. It is because of CEEC's nodes deployment planning and centralized clustering formation. In Fig .4 dead nodes with respect to rounds are described. Like earlier case, in CEEC perform better, and last node dies after almost 4200 rounds. In Fig 5 packet to BS is calculated for all routing protocols. It also shows that CEEC is very efficient in successful data delivery. Guaranteed number of CHs throughout the network operation improve the throughput of CEEC routing protocol. Fig. 6 shows the CHs selection of all routing protocols and is clearly understandable, how CEEC provides optimal number of CHs for every round. In Fig 6, numbers of selected cluster-heads per round are shown. From results it is clearly understandable that only CEEC is providing required number of CHs continuously. LEACH, SEP, E-SEP and DEEC do not provide guaranteed number of CHs per round. DEEC and SEP have more uncertainly in CHs selection in each round. Their uneven CHs generation, badly effects the amount of packets received by BS from CHs. Because of this CEEC has maximum throughput and network lifetime

\vspace{-0.3cm}
\begin{figure}[h]
\centering
\includegraphics[scale=0.45]{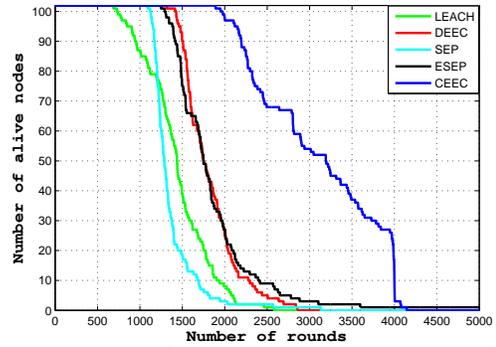}
\vspace{-0.5cm}
\caption{Alive Nodes for $100m\times100m$ Network with 100 nodes}\label{abc}
\end{figure}
\vspace{-0.3cm}

\vspace{-0.3cm}
\begin{figure}[t!]
\centering
\includegraphics[scale=0.45]{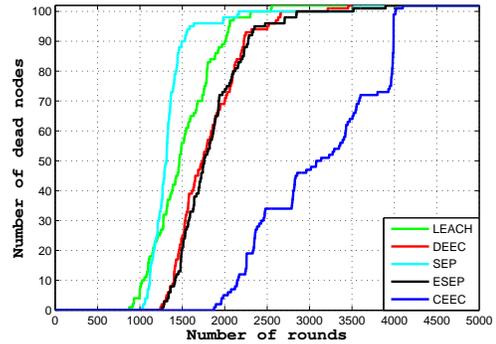}
\vspace{-0.5cm}
\caption{Dead Nodes for $100m\times100m$ Network with 100 nodes}\label{abc}
\end{figure}
\vspace{-0.3cm}

\vspace{-0.3cm}
\begin{figure}[t!]
\centering
\includegraphics[scale=0.45]{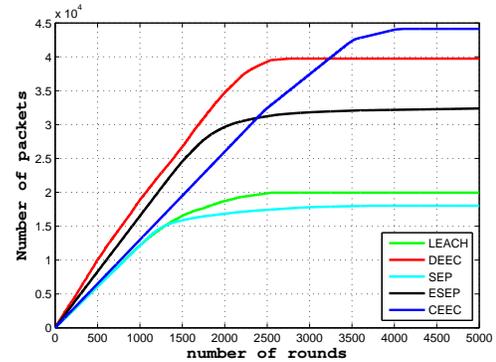}
\vspace{-0.5cm}
\caption{Packet to BS Nodes for $100m\times100m$ Network with 100 nodes}\label{abc}
\end{figure}
\vspace{-0.3cm}

\vspace{-0.3cm}
\begin{figure}[h]
\centering
\includegraphics[scale=0.45]{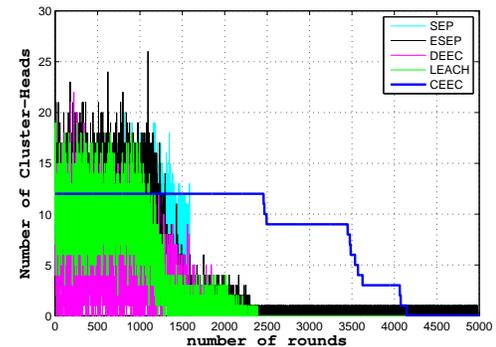}
\vspace{-0.5cm}
\caption{CHs per round}\label{abc}
\end{figure}
\vspace{-0.3cm}

\end{document}